\documentclass[a4paper,aps,prl,two column,10pt,floatfix]{revtex4-2}

\usepackage{graphicx}
\usepackage{xcolor}
\usepackage{tikz}
\usepackage{amsmath}
\usepackage{amssymb}
\usepackage{natbib}
\usepackage{notoccite}
\usepackage{color}
\usepackage{psfrag}
\usepackage{wasysym}
\usepackage{float}
\usepackage[normalem]{ulem}
\usepackage{pgfplots}
\usetikzlibrary{pgfplots.groupplots}
\usepackage{textcomp}
\usepackage[small,bf,justification=Justified]{caption}
\usepackage{subcaption}
\usepackage{ragged2e}
\usepackage{comment}

\pgfplotsset{compat=newest}
\pgfplotsset{yticklabel style={text width=1.3em,align=right}, xticklabel style={text width=1.5em,align=center}}
\pgfplotsset{every axis legend/.append style={anchor=south east}}

\begin{document}
	\title{Compliance-Shifting Elastogranular Shells}
	\author{Jairaj Narendran$^1$, Paolo Limcaoco$^1$, Shashank S. Basty$^1$, and Douglas P. Holmes$^1$}
	\affiliation{\footnotesize$^1$ Mechanical Engineering, Boston University, Boston, MA, 02215, USA}

	\date{October 31, 2026}
	
	\begin{abstract}
		Shells and grains separately exhibit complex nonlinear behavior when confined - shells buckle and snap, while grains transition from fluid-like to solid-like. When coupled, elastogranular structures have the ability to display a remarkable variety of mechanical behaviors. Here we created an elastogranular shell where spherical grains are sandwiched between two thin, elastic shells. By evacuating the air enclosed by the two layers, we can tune the stiffness of the elastogranular shell and alter its critical buckling load. Introducing controlled vacancy defects in the arrangement of the grains allows us to more finely tune critical buckling parameters, as the shell must balance the combined effects of defect size and resultant granular packing density. Defects in the jammed granular layer have a counterintuitive effect on the shell’s buckling load, as they can stiffen the shell away from the defect. With an empirically determined scaling model, we established how the shell and grains distribute load and create a compliance-shifting structure with a tunable buckling load.
	\end{abstract}
	
	
	\maketitle
	
	\section{Introduction}
	
	Shell structures are found in nature across many length scales such as in viral capsids~\citep{caspar_physical_1962,zandi_origin_2004}, pollen grains~\citep{katifori_foldable_2010,niklas_aerodynamics_1985}, and plant leaves~\citep{steele_shell_2000,moulia_leaves_2000}. Shell structures can withstand extreme loading conditions because they evenly distribute applied stresses along their curved middle surface. Their efficient load distribution, high strength-weight ratio, and unique appearance make them an attractive choice for engineered structures~\citep{huzel_modern_1992,jawad_structural_2018}. However, shell structures also display severe sensitivity to imperfections, which result in shells buckling at lower loads than predicted~\cite{koiter_effect_1963,hutchinson_imperfection_1967,koiter_stability_1967,hutchinson_postbuckling_1970}. The ratio of the actual buckling load to the theoretical value is referred to as the {\em knockdown factor}~\citep{zoelly_ueber_1915, von_karman_buckling_1939}. As buckling can be a catastrophic event in an engineered shell structure, the prediction of the critical buckling load on spherical shells has been an important and widely studied problem for over 80 years~\citep{tsien_theory_1942,huang_unsymmetrical_1964,koiter_nonlinear_1966,bushnell_buckling_1981,wunderlich_buckling_2002,pan_overview_2010,jimenez_technical_2017,nayyeri_amiri_nondestructive_2017,zhang_buckling_2017,marthelot_buckling_2017} Due to advances in fabricating soft elastic structures, spherical shells of fewer or more controlled defects can be fabricated, ushering in a revival of studying the impact of imperfections on shell characteristics~\citep{hutchinson_buckling_2016,lee_geometric_2016,lee_evolution_2019,holmes_nonlinear_2020}. However, shell parameters are locked into place during fabrication, which we see as an avenue of exploration.
	
	\begin{figure}[htbp]
		\centering
		\includegraphics[width=\linewidth]{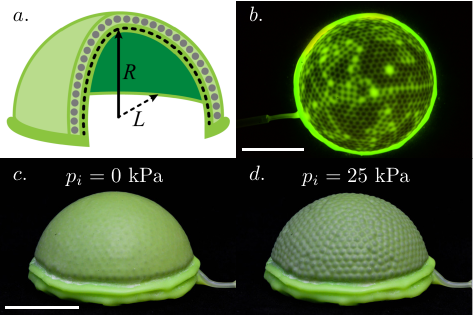}
		\caption{ 
			(a) A cutaway schematic of an EG shell. The middle surface of the EG shell is depicted using a dashed line on the inner shell; this is the point to which the radius $R$ and span $L$ of the EG shell is measured.
			(b) A backlit image of an EG shell with $\phi_0=0.74$.
			The same EG shell at (c) $p_i=0$ kPa and at (d) $p_i=25$ kPa. The EG shell at $p_i=25$ kPa has a bumpy outer surface because the applied negative internal pressure results in the outer shell conforming to the grains in the inter-shell cavity. All scale bars represents 30 mm. 
		}
		\label{shellOverviewPics}
	\end{figure}
	
	The ability to tunably change the stiffness or critical buckling load would allow for shells to be more widely used in metamaterial systems. Lee et al. created imperfect bilayer shells composed of two different crosslinked polymers capable of exchanging free molecular chains. Depending on the  orientation of the shrinking and swelling layers as well as the extent of differential swelling, the elastomeric shells exhibit a time-dependent increase or decrease of buckling pressure from its initial state - that is, a modification of the shell's knockdown factor~\citep{lee_evolution_2019}. Similarly, Yan et al. designed shells made of magnetic elastomer for which the buckling strength and knockdown factor could be dynamically tuned by changing the density and direction of the applied magnetic flux~\citep{yan_buckling_2020}.
	
	Elastogranular systems consist of a thin elastic component and a granular component. The jamming of the granular arrangement through the application of compressive forces allows for the system to switch between a fluid-like and a solid-like state. The soft, fluid-like state will rearrange and conform to different confining geometries, while the rigid, solid-like state preserves its shape after a jamming transition. Elastogranular interactions have been identified and studied in various natural phenomena such as in plant root growth mechanics~\citep{wendell_experimental_2012,kolb_radial_2012} and biological locomotion~\citep{maladen_undulatory_2009,young_vertical_2003}. Recent work has attempted to understand the mechanics of model elastogranular systems to apply them new engineered structures. Previous work has shown that large deformations of a confined slender elastic beam can result in jamming behavior in the surrounding grains due to increased stress concentrations overpowering the local ordering of the grains~\citep{schunter_elastogranular_2018,schunter_elastogranularity_2019}. In fact, many granular systems are functionalized by introducing elastic components which apply geometric constraints. Elastogranular systems have been used to create reinforcing and structural elements like exoskeletons~\citep{thompson-bean_soft_2015}, jammed architecture~\citep{aejmelaeus-lindstrom_jammed_2016,guerra_emergence_2021,svatos-raznjevic_jammed_2024}, and aerostructures~\citep{brigido-gonzalez_switchable_2019}. Various researchers have utilized elastogranular jamming via an applied negative pressure to create robotic grippers and manipulators capable of grasping objects of irregular geometries~\citep{brown_universal_2010,cheng_design_2012,joseph_jamming_2023,liu_support-enhanced_2026}. Guerra et al. varied the adhesion between grains and elastic sheets to explore its effects on sheet-grain conformality~\citep{guerra_elastogranular_2023}. They found that increasing adhesion between grains and a sheet in an elastogranular system increases the conformality of the sheet to the individual grains, which can restrict grain motion and result in a global change in shape.
	
	\begin{figure}[htbp]
		\centering
		\includegraphics[width=\linewidth]{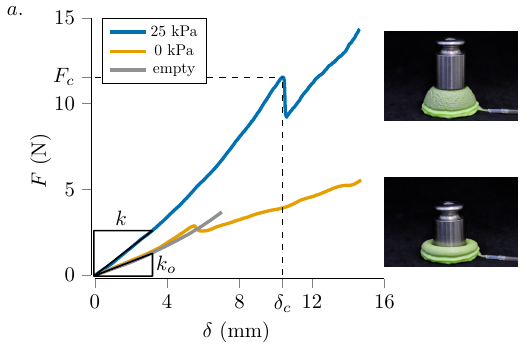}
		\includegraphics[width=\linewidth]{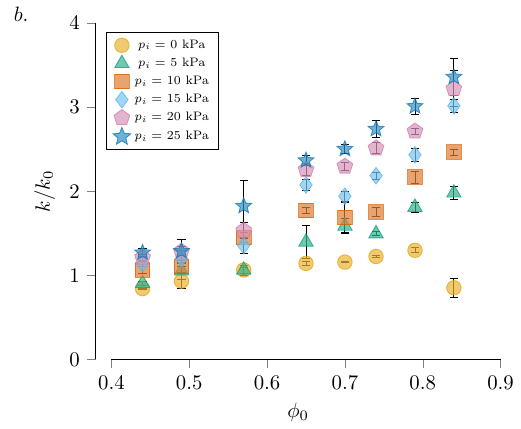}
		\caption{ 
			(a) The force-displacement response of compressing an EG shell with a packing fraction $\phi_0 = 0.74$ at internal pressures of $p_i = 0$ kPa and 25 kPa, as well as an empty EG shell ($\phi_0 = 0$), which we consider a classical shell, are plotted. The critical buckling force $F_c$, the critical buckling compression $\delta_c$, and the stiffness $k$ of the EG shell at $p_i = 25$ kPa are marked on the plot. The stiffness of an empty EG sell is denoted by $k_0$. The top and bottom inset images show the EG shell with $p_i = 25$ kPa and $p_i = 0$ kPa, respectively, when subject to a 500 g mass.
			(b) As the magnitude of the internal pressure $p_i$ is increased, the normalized stiffness $k/k_0$ also increases for increasing values of $\phi_0$ tested. The drop in stiffness at $\phi_0 = 0.84$ at $p_i=0$ kPa is due to the inner shell being slightly deformed by the mass of grains without any internal pressure to reinforce it, {\em i.e.} the weight of the grains is no longer insignificant.
		}
		\label{shellIntro}
	\end{figure}
	
	We took inspiration from phase change materials and elastic instabilities to apply granular jamming to spherical shells to make structures with compliance-shifting capabilities (Fig.~\ref{shellOverviewPics}). To create a shell with tunable mechanical properties, we interleaved a monolayer of grains (with a radius $r_g$) between two elastic shells (an thicker inner shell of radius $h_i$ and a thinner outer shell of radius $h_o$) and sealed this {\em elastogranular} (EG) {\em shell} along its base. We defined the total thickness of the EG shell as $h_t = h_i + 2r_g + h_o$. The resulting EG shell has a span $L$ and radius $R$, but because the shell is approximately hemispherical, we considered $L \approx R$. 
	
	The term $p_i$ represents the magnitude of the negative pressure applied to the inter-shell cavity, which is the sealed region between the inner and outer shells. For example, a $p_i = 0$ kPa refers to the inter-shell cavity as ambient pressure, while ``increasing'' the internal pressure to $p_i = 25$ kPa represents applying $-25$ kPa to the inter-shell cavity. The EG shell at these two states are displayed in Fig.~\ref{shellOverviewPics}c-d.
	
	The concave side of the EG shell is open to ambient pressure and is equivalent to the pressure on the convex side of the EG shell: both are at $p_{atm}$. Whereas Guerra et al. achieved increased sheet-grain conformality using a magnetic sheet and grains, the EG shell uses an applied negative internal pressure. In Fig.~\ref{shellOverviewPics}c, the EG shell is at $p_i = 0$ kPa and the outer shell is not conformed to the grains while in Fig.~\ref{shellOverviewPics}d, the EG shell is at $p_i = 25$ kPa and the outer shell conforms to the enclosed grains. A more detailed explanation of EG shell construction is described in the Appendix ~(Fig.~\ref{shellFab}).
	
	In this work, we explored the EG shell characteristics at different deflection regimes. We begin by varying the inter-shell cavity pressure and enclosed granular packing density to understand their effects on the stiffness of the EG shell in the small deflection regime ({\em i.e.} $\delta<h_t$). Additionally, we described possible mechanisms for the compliance-shifting behavior observed. Building upon this knowledge, we used the Buckingham $\Pi$ theorem to identify the relevant shell parameters and derive scaling relations to characterize compression force in these new EG shells based on existing theory for monolayer elastic shells. Next, we applied larger deflections to EG shells to explore their critical buckling characteristics in the large deflection regime ({\em i.e.} $\delta>h_t$). We introduced granular vacancy defects as the primary, first order imperfection with the greatest effect on the load bearing capacity of EG shells. A combination of different defect locations and indentation locations are tested to understand the local and global effect of defects, as well as how EG shells react to eccentric loading. Finally, the transformation of the enclosed grains from fluid-like to solid-like upon the application of a negative differential pressure is utilized to explore a shape-locking mechanism in EG shells which can also yield varying stiffness responses.
	
	\section{Small Deflection Regime}
	
	\subsection{Compliance-Shifting}
	
	For a shell with a given initial granular packing density ($\phi_0$), we varied $p_i$ and measured the stiffness of the resulting shell. We measured the stiffness by performing uniaxial compression tests with a plate load applied to the apex of the shell~Fig.~\ref{shellParameters}c. A representative plot of resulting force vs displacement curves for an EG shell with $\phi_0 = 0.74$ is shown in Fig.~\ref{shellIntro}a. For this shell, we tested two values of $p_i$: $p_i=0$ kPa (yellow) and $p_i=25$ kPa (blue), and see that decreasing the internal pressure increases the EG shell's stiffness. When the deflection of the EG shell's apex is small relative to the shell's thickness ({\em i.e.} $\delta < h_t$), the shell behaves like a linear, elastic spring of stiffness $k$, such that $F = k\delta$. We see that that slope of the linear region when $p_i=$25 kPa is steeper than when $p_i=$0 kPa, and we refer to this behavior as {\em compliance-shifting}. The linear behavior is consistent with the mechanical response of classical elastic shells in the {\em small deflection regime~\citep{reissner_stresses_1946, vella_indentation_2011, lazarus_geometry-induced_2012, yang_peripheral_2016}.} The stiffness of a classical thin elastic shell in the small deflection limit ($\delta \ll h$) is dictated by its materials and geometry~\citep{reissner_stresses_1946,vella_indentation_2011}, $k=8B/\ell_b^2$, where $B$ is the bending rigidity $B=Eh^3/12(1-\nu^2)$, and the length of boundary layer is $\ell_b = (BR^2/Eh)^{1/4} \sim \sqrt{hR}$, such that 
	\begin{equation}
		\label{Kdimensional}
		k\approx \frac{2}{3}\frac{E h^2}{R(1-\nu^2)}.
	\end{equation}
	Although the parameters $E$, $h$, and $R$ were unchanged in~Fig.~\ref{shellIntro}a, we observed an increase in shell stiffness. In the limit of no grains or no internal pressure, we expect these shells to behave classically. This motivates modifying the stiffness to account for the {\em effective} material and geometric properties that emerge due to the jamming granular layer. To provide a basis of comparison, we denote the stiffness of a classical shell with no grains ({\em i.e.} $\phi_0=0$) as $k_0$, and we will scale $k$ by $k_0$ to quantify the magnitude of the compliance-shifting. An alternate choice for $k_0$ would be the stiffness of the elastogranular shell ({\em i.e.} $\phi_0\neq 0$) when $p_i=0$, and looking at~Fig.~\ref{shellIntro}a we see that these values are quite similar. We note that the $h_i$ or the thickness of the inner shell is the same in both cases.
	
	\begin{figure}[htbp]
		\centering
		\includegraphics[width=0.95\linewidth]{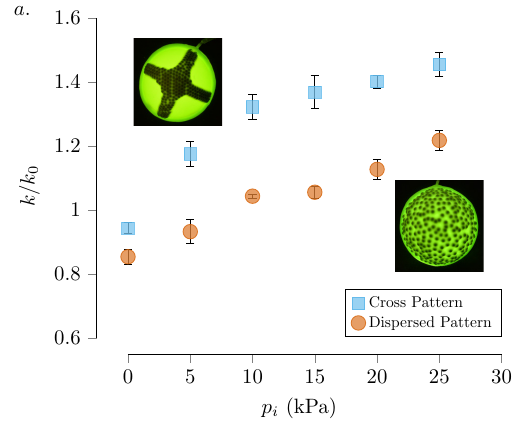}
		\includegraphics[width=0.95\linewidth]{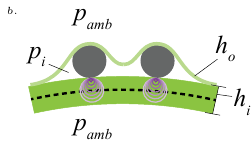}
		\includegraphics[width=0.95\linewidth]{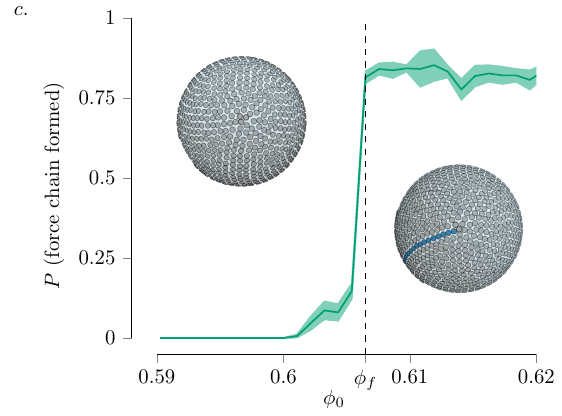}
		\caption{ 
			(a) An EG shell of $\phi_0 = 0.28$ arranged into a cross pattern can achieve higher values of normalized stiffness $k/k_0$ than in a dispersed pattern, although both shells show increased stiffness with increasing $p_i$.
			(b) Negative pressure causes the outer shell to conform to the grains, causing a Hertzian pressure distribution (purple circles) which prestresses the inner shell. The middle surface of the EG shell is denoted with the dashed line.
			(c) Simulation results showing the probability of developing a force chain as a function of $\phi_0$. More than half of our trials yielded force chains when $\phi_0 = \phi_f = 0.606$. Inset are visualization results of a dome with $\phi_0=0.59$ on the left and a dome with $\phi_0=0.62$ on the right with a successful force chain highlighted.
		}
		\label{forceChainMechanism}
	\end{figure}
	
	To see how the compliance-shifting is affected by the initial grain packing density, we tested shells with different $\phi_0$ and plot $k/k_0$ as a function of $\phi_0$ at different values of $p_i$ in~Fig.~\ref{shellIntro}b. As $\phi_0$ is increased from 0.44 to 0.84 for the same $p_i$, there is a general linear trend of increased stiffness for a given value of $p_i$. The stiffness of shells with $\phi_0 <$ 0.65 show an increase in stiffness that less than two times the stiffness of an empty EG shell. EG shells with $\phi \geq 0.65$ show an increase in stiffness that is more than twice that of an empty EG shell and additionally exhibits a larger range of stiffness that can be achieved by tuning $p_i$. An EG shell with $\phi_0 =$ 0.84 shows a normalized stiffness $1 < k/k_0 < 3$ at 25 kPa; showcasing it's tunable compliance-shifting ability.
	
	An EG shell with $\phi_0 = 0.28$ was fabricated and the grains were organized into two conformations to test their effect on stiffness response in Fig.~\ref{forceChainMechanism}a. In the EG shell with a cross pattern (left inset image), the grains are arranged to be in contact with each other. In the EG shell with a dispersed pattern (right inset image), the grains were evenly spaced to minimize contact between neighboring grains.

	Our first observation is that the EG shells exhibit an increase in stiffness $k/k_0$ with increasing $p_i$ regardless of pattern and while $\phi_0$ remains constant. When a negative pressure ($p_i >$ 0) is applied to the EG shell, the outer shell has a greater effective adhesion to the enclosed grains and starts to conform to them. As $p_i$ is increased, this \emph{conformality} -- as defined by Guerra et al. in elastogranular sheets~\citep{guerra_elastogranular_2023} -- of the shell to the grains also increases (see~Fig.~\ref{conformalityExp}b) and pushes the grains into the inner shell. Higher values of $p_i$ therefore increasingly prestress the inner shell, resulting in an increased stiffness response upon shell compression. As long as there are grains present for which the conformality can be changed by tuning $p_i$, this prestress mechanism is present in EG shells. A schematic of this prestress mechanism is depicted in Fig.~\ref{forceChainMechanism}b, with the purple circles representing the Hertzian pressure distribution.
	
	We also see that the EG shell with the cross pattern exhibits a higher stiffness response $k/k_0$ than the EG shell in the dispersed pattern despite both shells having the same $\phi_0$. This suggests that the increased stiffness is a product of granular arrangement, specifically the grains in contact. Applying a load to a contiguous granular media results in the formation of a network of force chains between the individual grains to support the load and propagate the stresses~\citep{cates_jamming_1999}. We theorize that these force chain networks can also develop in EG shells if the grains are in contact, and that having force chains that propagate from the apex to the edge of the shell allow it to exhibit a greater stiffness response.
	
	In Fig.~\ref{forceChainMechanism}a, we have shown that at low packing fractions (i.e. $\phi_0 = 0.28$) evenly spaced grains cannot develop force chains. We used a Monte Carlo simulation to place grains on a spherical dome and find the minimum packing fraction $\phi_0$ necessary for force chain development. More details about the simulation can be found in Appendix A.3. In Fig.~\ref{forceChainMechanism}c, we plot simulation results for the average probability of force chains being formed for 5 sets of 100 trials at each value of $\phi_0$. For our simulations where $\phi_0 < 0.6$, less than 5\% of the trials yielded a force chain spanning from the apex to the edge of the shell. In all of our simulations where $\phi_0 \geq 0.61$, we observed at least one force chain in over 70\% of trials. Visualization of the domes at $\phi_0 = 0.59$ and $\phi_0 = 0.62$ are shown as the left and right inset images, respectively, in~Fig.~\ref{forceChainMechanism}c, with the individual force chains highlighted. We find that in our simulations, the packing fraction where more than half of the trials yield force chains is at $\phi_0=0.606$, which we denote as the cutoff for force chain development $\phi_f$; this value is also marked in~Fig.~\ref{shellIntro}c. As the simulations show, this cutoff is not definitive and shells with $0.6 \leq \phi_0 \leq 0.606$ may also develop force chains. Therefore increased stiffness behavior for the shells we tested with $\phi_0 \geq 0.606$ can thereby be attributed to force chain propagation in addition to a prestressed shell.
	
	\begin{figure}[htbp]
		\centering
		\includegraphics[width=0.95\linewidth]{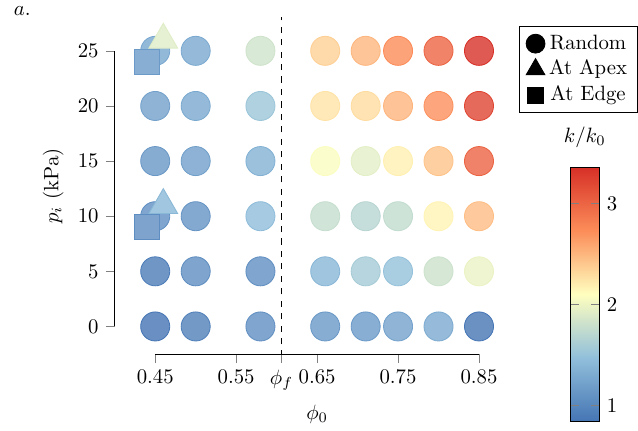}
		\includegraphics[width=0.95\linewidth]{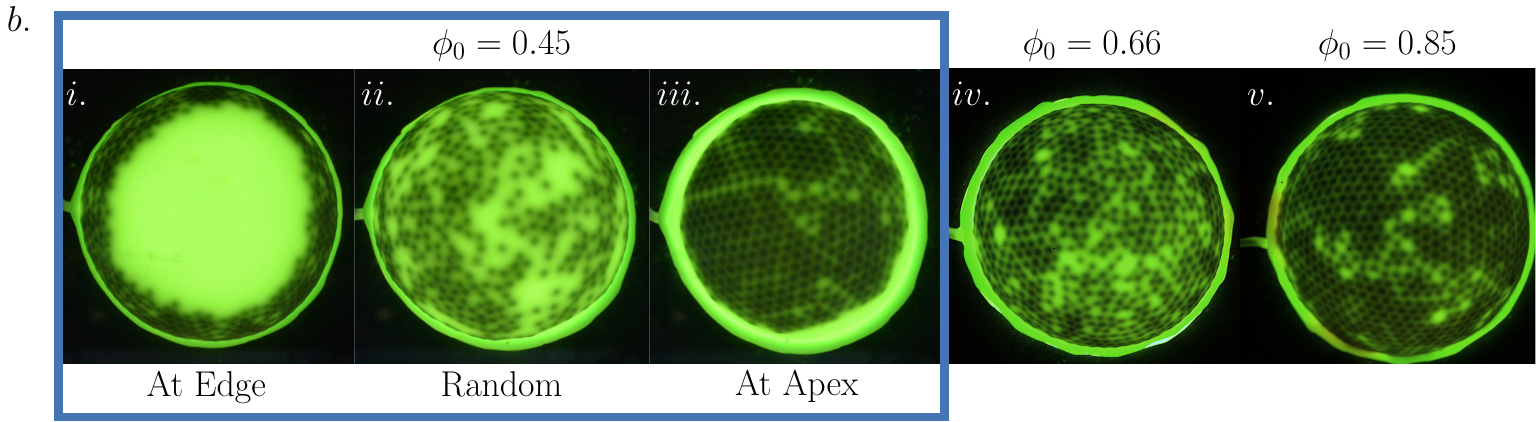}
		\caption{ 
			(a) The stiffness $k/k_0$ for all tested combinations of $p_i$ and $\phi_0$. We denote the \emph{elastogranular} regime as shells with $p_i > 0$ kPa and $\phi_0 > \phi_f$. Shells with $\phi_0 = 0.45$ at $p_i = 10$ kPa and 15 kPa are measured for stiffness with different granular arrangements. 
			(b) Images of EG shells at different values of $\phi_0$. The three granular arrangements tested for stiffness for the shell with $\phi_0 = 0.45$; the grains are rearranged to the (i) edge, (ii) randomly, or (ii) apex to understand the effect of granular arrangement on shell stiffness behavior. (iv-v) shows EG shells at $\phi_0 = 0.66$ and 0.85. Notice fewer vacancies and more regions of continuous hexagonal packing in the $\phi_0=0.85$ case.
		}
		\label{allDataColormap}
	\end{figure}
	
	The effect of different combinations of $p_i$ and $\phi_0$ on normalized stiffness $k/k_0$ is plotted in~Fig.~\ref{allDataColormap}a. To reiterate the importance of granular arrangement on EG shell stiffness, a shell with $\phi_0 = 0.44$ was arranged into edge, evenly dispersed, or center configurations~Fig.~\ref{allDataColormap}b i-iii and then tested for stiffness response. As shown for this EG shell at $p_i = 10$ kPa and 25 kPa, a shell with the same $p_i$ and $\phi_0$ can exhibit different values of stiffness depending on the granular arrangement. In the edge configurations, the indenter pushes onto a point where the local $\phi = 0$ and prestressing of the shell is only at the edges, away from indentation. In the center configurations, partial force chains can form at the apex of the shell, increasing the stiffness response. All the other data is from shells in the dispersed configuration. Fig.\ref{allDataColormap}b iv-v are representative images of shells with $\phi_0 =$ 0.65 and 0.84. 
	
	The physical responses of the EG shells in~Fig.~\ref{allDataColormap} a to small compressive displacements can be classified into three sets of behaviors motivated by the separate mechanisms mentioned above. At $p_i = 0$ kPa, we observed that EG shells of all values of $\phi_0$ exhibit stiffness $k/k_0 < 1.5$. For this group of EG shells, the outer shell does not conform to the enclosed grains. Therefore, the inner shell is not prestressed and force chains cannot develop when the EG shell is subject to a load because the grains are not locked into place. For EG shells with $p_i \geq 5$ kPa and $\phi_0 < \phi_f$, the inner shell is prestressed by the increased conformality of the outer shell to the grains, but the individual grains are too far apart for force chains to develop. For shells with $\phi_0 > \phi_f$ and $p_i \geq 5$ kPa, both inner shell prestress as well as force chain development allows for increased stiffness behavior. This third group of shells displays both conformality of elastic sheets and force chain development across the granular layer, and we will refer to it as the \emph{elastogranular compliance-shifting regime}. For further investigation into understanding the scaling relations for our EG shell as well as the effect of vacancy defects, we will focus on data in this elastogranular compliance-shifting regime. 
	
	In bulk granular media, there are jammed neighboring grains in the direction of the applied load which keep the initial grains that the load contacts in place so that force chains can develop. In EG shells, we believe that the thicker inner shell plays the role of the bulk granular media in supporting the enclosed monolayer of grains so that force chains may develop. If we were to switch the thickness of the inner and outer shells, it would be the inner shell conforming to the grains when the EG shell is subject to a negative pressure. It would be unclear if force chains would be able to form in this case as there are no stiffer layer or bulk media in the direction of the applied load past the monolayer of grains.
	
	\subsection{Effective Compression Force}
	
	For small displacements ($\delta \ll h$), the force $F$ linearly increases with displacement $\delta$ proportional to a spring whose stiffness is given by eq.~\ref{Kdimensional}. To compare the stiffness of different shells, we can non-dimensionalize this linear response -- for example, by scaling the Donnell-Mushtari-Vlasov  (DMV) equation -- to find that~\citep{vella_indentation_2011, taffetani_static_2018}
	\begin{equation}
		\label{nondimFvD}
		\bar{F} = \frac{8}{\lambda^2}\bar{\Delta},
	\end{equation}
	where 
	\begin{equation}
		\label{dimlessF}
		\bar{F} = \frac{FL^3}{R^4Eh}\approx\frac{F}{REh},
	\end{equation}
	and 
	\begin{equation}
		\label{dimlessD}
		\bar{\Delta} = \frac{\delta R}{L^2}\approx\frac{\delta}{R},
	\end{equation}
	where we note that $R \approx L$ for the EG shells studied in this paper. The prefactor $\lambda$ in eq.~\ref{nondimFvD} emerges from a balance of the shell's stretching energy to its bending energy; the fourth root of this balance is commonly used~\citep{libai_nonlinear_1998, brinkmeyer_pseudo-bistable_2012,taffetani_static_2018}, such that, 
	\begin{equation}
		\label{lambda}
		\lambda = [12(1-\nu^2)]^{1/4}\sqrt{\frac{R}{h}}\alpha,
	\end{equation}
	where $\alpha$ is the angle subtended from the pole to the edge of the shell. For the approximately hemispherically shells studied in this paper, $\alpha=\pi/2$.
	This classical relation in eq.~\ref{dimlessF} is able to capture EG shell behavior at $p_i = 0$ kPa, but does not account for the role of the grains and the inter-shell pressure. In this section, we will use a scaling analysis to determine how these additional parameters contribute to the stiffness of an elastogranular shell.
	
	To proceed, we used the Buckingham $\Pi$ theorem to determine the relevant parameters to characterize compression force. Since our experimental results confirm a linear dependence of force on displacement for our elastogranular shells~\ref{shellIntro}a, we seek to modify eq.~\ref{nondimFvD} with an effective force that accounts for the components of the EG shell, such that
	\begin{equation}
		\label{EGnondimFvD}
		\bar{F}_{eg} = \frac{8}{\lambda^2}\bar{\Delta}.
	\end{equation}
	We rewrite the dimensionless force given by eq.~\ref{dimlessF}, in terms of several dimensionless $\Pi$ groups,
	\begin{equation}
		\bar{F}_{eg} = \bar{F}\cdot f\left(\frac{E}{p_i}, \frac{h_o}{h_i}, \frac{r_g}{R},\frac{\phi_0}{\phi_f}\right)
	\end{equation}
	where
	\begin{equation}
		\Pi_1 \equiv \frac{E}{p_i}, \\\ \Pi_2 \equiv \frac{h_0}{h_1}, \\\ \Pi_2 \equiv \frac{r_g}{R}=, \\\ \Pi_4 \equiv \frac{\phi_0}{\phi_f}.
	\end{equation}
	
	In the previous section, we identified the shells in the elastogranular regime as the focus of our investigation. Next, we created EG shells that vary the different geometric and material parameters found in each $\Pi$ group while keeping the other parameters constant, and test their force response up to $\delta = R/10$ for different values of $p_i$. For $\Pi_1$, we tested shells made of three different Young's modulus $E$. For $\Pi_2$, we test shells with three different inner shell thickness $h_i$. For $\Pi_3$, we tested two different shell radii $R$ as well as two different grain radii $r_g$. For $\Pi_4$, we tested four different packing fractions $\phi_0$ in the elastogranular regime.
	
	\begin{figure}[htbp]
		\centering
		\includegraphics[width=0.95\linewidth]{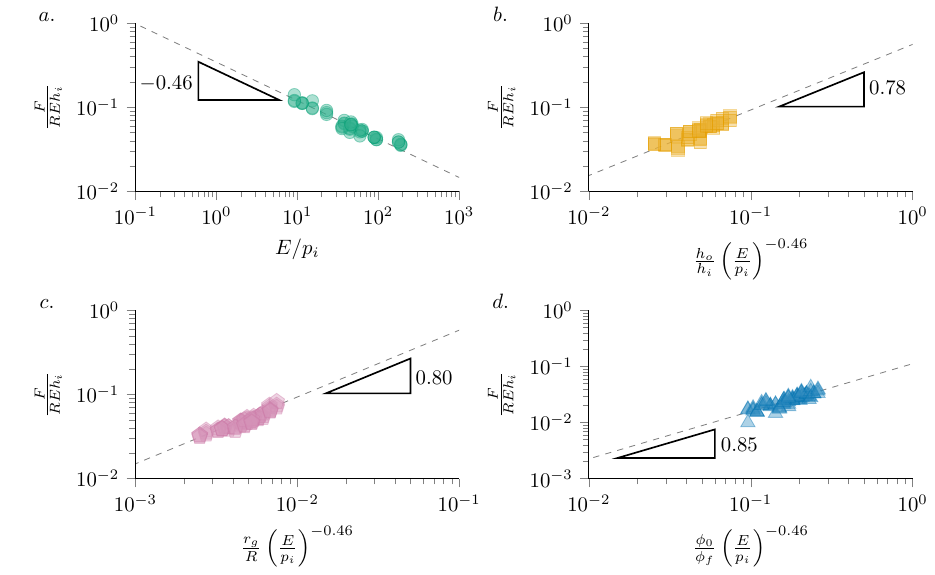}
		\includegraphics[width=0.95\linewidth]{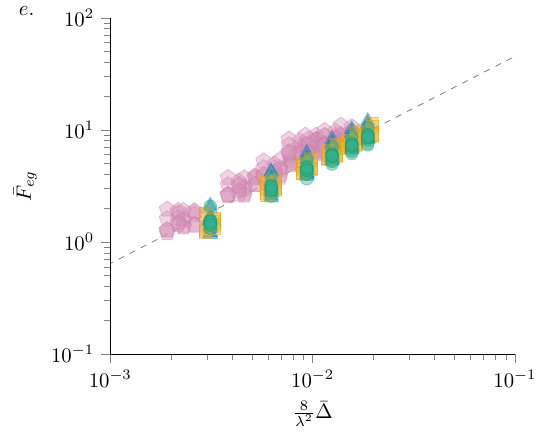}
		\caption{
			EG shell data is plotted by their $\Pi$ groups against classical dimensionless force. The $\pi$ groups include (a) $E/p_i$ (b) $h_o/h_i$ (c) $r_g/R$ and (d) $\phi_0/\phi_f$. The scaling for $E/p_i$ is added on to the other three $\pi$ groups here to accurately compare EG shell data of multiple $p_i$ to get more data for scaling analysis. (e) All EG shell data in the elastogranular regime is collapsed using eq.~\ref{EGnondimFvD} with the scaling from eq.~\ref{new_scaling}.
		}
		\label{scalingPlots}
	\end{figure}
	
	Each $\Pi$ group is plotted on a log-log scale against their dimensionless force $F/(REh_i)$ for the different shells in the elastogranular regime in~Fig.~\ref{scalingPlots}a-d. We first plot the $\Pi_1$ and use a linear regression to find its scaling relation against dimensionless force as $-0.46 \pm 0.03$ within a 95\% confidence interval. For plots b-d, $\Pi_2-\Pi_4$ are multiplied by the scaling for $\Pi_1$ so that data points from different values of $p_i$ can be compared together, i.e. we tested three values of $h_o/h_i$ at five values of $p_i$ so that we can calculate the fit with fifteen data points. We then used linear regression on the data to derive the scaling values for $\Pi_2-\Pi_4$ groups. The R$^2 > 0.8$ for the linear regressions for all four plots. After acquiring the scaling relations, we can rewrite the new dimensionless force with the $\Pi$ groups, which represent extra terms present within to EG shells as
	\begin{equation}
		\label{new_scaling}
		\bar{F}_{eg} = \frac{F}{REh_i}
		\left(\frac{E}{p_i}\right)^m
		\left(\frac{h_o}{h_i}\right)^n
		\left(\frac{r_g}{R}\right)^p
		\left(\frac{\phi_0}{\phi_f}\right)^q
	\end{equation}
	where $m = 0.46 \pm 0.03$, $n = -0.78 \pm 0.12$, $p = -0.80 \pm 0.08$, and $q = -0.85 \pm 0.11$, all within a 95\% confidence interval. It should be noted that the scaling relations are negative compared to their results in the linear regressions because each $\Pi$ group is being plotted with dimensionless force $\bar{F}_{eg}$ as in eq.~\ref{new_scaling} instead of against dimensionless force $F/(REh_i)$ as in Fig.~\ref{scalingPlots}a-d. To validate that this scaling matches our data, we plot the new derived dimensionless force $\bar{F}_{eg}$ as in eq.~\ref{EGnondimFvD} using force data taken at $\delta \in$ [0.5, 1, 1.5, 2, 2.5, 3] mm for shells in the elastogranular regime in~Fig.~\ref{scalingPlots}d.We found that the data collapses in the small deflection regime with an R$^2 = 0.87$.
	
	We realize there are limitations to the scaling we have derived. The scaling relation for $\Pi_2-\Pi_4$ groups assumes that the scaling for $\Pi_1$ is correct. We varied $\Pi_1$ over a decade and find an R$^2 = 0.95$, but testing a larger range of values would allow us to be more confident. We varied the other three $\Pi$ groups over half a decade, which allows us to capture the behavior of shells within the range of geometries we tested. Future work where two or three decades of data for each $\Pi$ group are explored could allow us to apply this scaling to EG shells with a wider range of geometric properties. 
	
	\section{Large Deflection Regime}
	
	\subsection{Controlled Buckling Behavior}
	
\begin{figure}[htbp]
		\centering
		\includegraphics[width=0.95\linewidth]{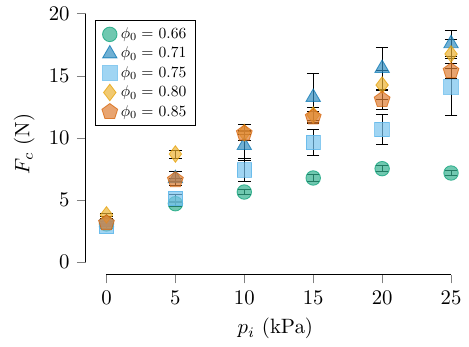}
		\caption{
			For EG shells in the elastogranular regime, $F_c$ increases with $p_i$, but the EG shell with the higher $\phi_0$ does not necessarily exhibit the highest $F_c$. The EG shell with $\phi_0 = 0.66$ may have distinctly lower values of $F_c$ due to fewer force chains.
		} \label{FcPhi}
	\end{figure}

	In Fig.~\ref{shellIntro}a, we observed distinct values of $\delta$ where the force response of the EG shell suddenly decreases in both the 0 kPa and 25 kPa cases. We denoted this event for the EG shell at 25 kPa with $\delta_c$ and $F_c$, which refer to the critical buckling force and critical buckling compression distance, respectively. The critical buckling force of the EG shell at $p_i = 25$ kPa is almost a five-fold increase compared to at $p_i = 0$ kPa for the EG shell with the same $\phi_0$. The displacement at critical buckling $\delta_c$ is a two fold increase between the two values of internal pressure as well. In Fig.~\ref{FcPhi}, we plot the effect of $p_i$ on $F_c$ for a range of $\phi_0$ that are in the elastogranular regime. There is a general trend of increasing $F_c$ as $p_i$ is increased. However, when compared to each other at the same $p_i$, larger values of $\phi_0$ do not necessarily result in larger values of $F_c$. The exception to this case being $\phi_0 = 0.66$, which displays much lower values of $F_c$ due to fewer force chains present, which has been observed in other systems~\citep{guerra_elastogranular_2022}. Additionally, the large standard deviations for $F_c$ point to a large variability in the experimental buckling load for an EG shell with a specific $\phi_0$ and $p_i$, which we have not explored up till this point. 
	
	Buckling refers to the phenomena in shells where the structure deforms nonlinearly when an applied load reaches a critical point~\citep{hutchinson_imperfection_1967}. In our EG shells, the critical buckling load event is denoted by a sudden force drop in the force vs compression plot. We believe that the number and arrangement of these grains -- which serve as the middle surface of the shell -- are what dictate the range of buckling properties in the EG shell. As shown in Fig.~\ref{conformalityExp}b, higher values of $p_i$ will increase the conformality of the outer shell to the enclosed grains. This makes it harder for the grains to rearrange and break the force chains, allowing the EG shell to resist a higher load before collapse. But force chains are only one part of the physical picture when it comes to the buckling of EG shells. With classical shells, variations in key shell characteristics, such as the critical buckling load, are attributed to defects in the middle surface of the shell~\citep{hutchinson_buckling_2016}. Gaussian dimples are the most commonly studied defects in thin shells, and are characterized by a small, axisymmetric depression without thickness variation or residual stresses~\citep{hutchinson_buckling_2016,abbasi_comparing_2023}. A less commonly studied defect is the through thickness defect, which refers to a region on the shell surface which is a ``soft spot'' where the thickness is less than the rest of the shell~\citep{yan_controlled_2016,paulose_buckling_2013}. In the case of both of these defect types, buckling behavior is dominated by the largest defect in the shell and instability occurs well below the theoretical values is observed~\citep{derveni_most_2025}.
	
	In our experiments, the initial compression of an EG shell in the elastogranular regime results in the formation of force chains in the enclosed grains that resist indentation. As the indentation continues, the cap of the EG shell is flattened, and when the EG shell reaches a critical force value $F_c$, the flattened EG shell cap delaminates from the indenter and everts inwards. At this point, the flat plate indenter only makes contact with the shell cap at the circular rim of this concave conformation, which in the limit of $h \rightarrow 0$ is sometimes referred to as the Pogorelov ridge~\citep{pogorelov_bendings_1988}. These two initial phases of hemispherical shell compression by a flat plate have been described by Yang et al.~\citep{yang_peripheral_2016} and confirmed by previous work as well~\citep{nasto_localization_2013,vaziri_mechanics_2009,updike_axisymmetric_1970}. It is important to note that $F_c$ depends on both the intrinsic physical properties of the EG shell as well as the frictional interaction between the surfaces of the EG shell and the flat plate indenter. A greater coefficient of friction between the surface of the EG shell and the surface of the flat plate indenter will result in a larger load required to cause shell indentation, which is in agreement with observations in previous work~\citep{nasto_localization_2013}.
	
\begin{figure}[htbp]
		\centering
		\includegraphics[width=0.95\linewidth]{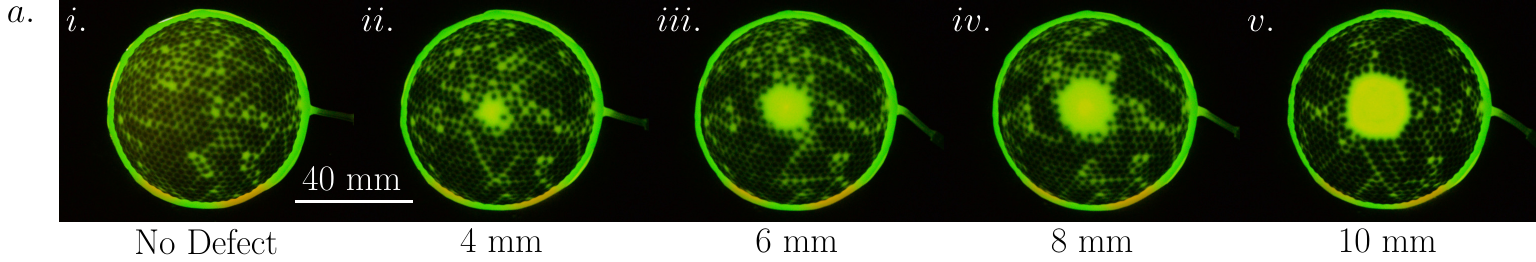}
		\includegraphics[width=0.95\linewidth]{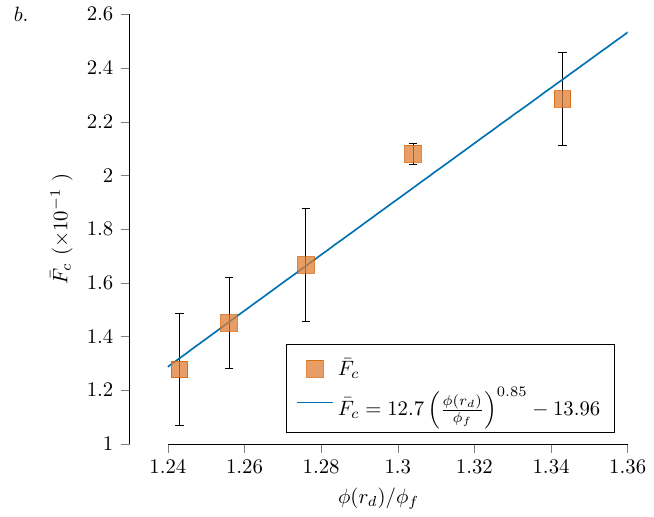}
		\caption{
			(a) Vacancy defects of varying sizes were made at the apex of an EG shell with $\phi_0 = 0.75$. 
			(b) On the x-axis is the packing fraction of the surrounding grains normalized by the critical packing fraction, denoted as $\phi(r_d)/\phi_f$ and on the y-axis is the dimensionless critical buckling force $F_c/(REh_i)$. The square marks refer to the actual data points for a shell with different sized defects while the line represents the relation for $\Pi_4$ from Fig.~\ref{scalingPlots}d. As packing fraction increases, the critical buckling force also increases even though initial packing fraction ($\phi_0=0.75$) and internal pressure ($p_i=5$ kPa) are kept constant.
		} \label{DefectIntro}
	\end{figure}

	The large standard deviations in buckling loads in Fig.~\ref{FcPhi} can be attributed to the different sized randomly-occurring vacancy defects that are present when packing spheres on a spherical surface. When we introduced grains in the inter-shell cavity, new types of defects emerged. Due to the geometric frustration of hexagonal packing on a curved surface of the shell, defects in the crystal structure in the form of grain boundaries and vacancies now appear.  Since current research suggests that the actual buckling load is governed by the largest defect~\citep{derveni_most_2025}, we introduced grain vacancies as the largest, first-order defect within an elastogranular shell. These grain vacancies are regions of a given size in the granular media where no grains are present. By rearranging the grains to control the size and location of this vacancy defect, we investigated how granular packing defects affect the buckling behavior of EG shells. We introduced vacancy defects of 4 mm $\leq r_d \leq$ 10 mm to the apex of an EG shell with $R = 29.5$ mm, thickness $h_{i} = 0.18$ mm, and $\phi_0 = 0.75$ with a $p_i =$ 5 kPa applied to keep the defect in place. Defect sizes 0 mm $< r_d <$ 4 mm were not tested due to naturally occurring vacancy defects of similar sizes that arise due to the frustration of hexagonally packing spheres on a curved surface. Introducing a vacancy defect of that size would not be guaranteed to be the largest defect on the EG shell. Images of these defects on the same shell are seen in Fig.~\ref{DefectIntro}a. The process for defect creation is explained in more detail in Fig.~\ref{packingDefectCreation}.
	
	The EG shell is uniaxially compressed by a flat plate indenter until $\delta$ = 15 mm and the force-displacement response is measured. This displacement distance was chosen because all the shells displayed buckling behavior within this compression distance. As the size of intentional defects that we added are larger than the inherent grain vacancies, their effect on the packing fraction of the surrounding grains is nontrivial. When there is a vacancy defect on the apex of the shell, the grains are forced to rearrange into the annular hemisphere around the defect. This annular hemisphere has a smaller area than the whole shell and therefore packing the grains in this area results in an increased packing fraction $\phi(r_d)$. 
	
	To calculate $\phi(r_d)$, we first found the half-angular width of the defect $\gamma_d$ (see Fig.~\ref{packingDefectCreation}) as
	\begin{equation}
		\gamma_d=\arcsin(r_d/R)
	\end{equation}
	which is then used to find the surface area of the spherical cap covered by the defect $A_d$ as
	\begin{equation}
		A_d=2\pi R^2 (1-\cos(\gamma_d))
	\end{equation}
	where $h \approx R$ in the equation for the area of a spherical cap. Now defining total surface 2-dimensional area of the grains $A_{gT}=n_g A_g$ from eq.~\ref{phi0calc} yields
	\begin{equation}
		\phi(r_d) = \phi_0^{max}\left(\frac{A_{gT}}{A_s-A_d}\right)
	\end{equation}
	where $\phi_0^{max} = 0.86$ is the maximum packing fraction of grains on a spherical shell and $A_s$ is the total surface area of the spherical cap, also defined in Appendix A.1. As Fig.~\ref{shellIntro}b has shown that increasing $\phi_0$, or overall packing fraction, increases the stiffness of the shell, increasing $\phi(r_d)$ may also result in increasing critical buckling force as well if the indenter is ``masking'' the defect. The region of the EG shell that has been flattened through indentation is parallel to the flat plate indenter. If there is an applied load, only the grains at the edge of the indented region can propagate the forces through force chains down the rest of the unindented EG shell. Larger values of $\phi(r_d)$ means that there is higher density of grains in the unindented region so more force chains can develop to resist the indentation force.
	
	We plot dimensionless force $\bar{F}_c$ as a function $\phi(r_d)/\phi_f$ which is the packing fraction of the surrounding grains in the presence of a defect normalized by critical packing fraction in Fig.~\ref{DefectIntro}b. We observed that $\bar{F}_c$ increases as defect size increases even though initial packing fraction $\phi_0$ and internal pressure $p_i$ were kept the same. Our scaling for the small displacement behavior, specifically for $\Pi_4$ where $\bar{F}_{eg}\propto({\phi_0/\phi_f})^{0.85}$, suggests that buckling load may also follow the same relation. We replaced $\phi_0$ in $\Pi_4$ with $\phi(r_d)$ to apply it to EG shells with defects and fit it to our experimental data with a linear regression to get the scaling equation
	\begin{equation}
		\bar{F_c}=12.7\left(\frac{\phi(r_d)}{\phi_f}\right)^{0.85}-13.96
	\end{equation}
	with an R$^2=0.97$. The good agreement between the data and the fit suggests that the scaling for $\bar{F}_{eg}$ also seems to predict buckling behavior. 
	
	Our masking hypothesis for increased $F_c$ naturally leads to investigating the effect of defect location when it is not coincident with the indentation location. Others have also explored this interaction in the past, albeit with a point indenter, and found that an indentation probe and a defect may only interact if they are local to each other within a characteristic length $\sqrt{Rh}$ along the surface of the shell~\citep{abbasi_probing_2021}. Similarly, we would like to investigate whether vacancy defect location and flat plate indentation location interact in EG shells. The indentation location is set to be perpendicular to the apex of the shell while the defect location is varied along the meridian of the shell. 
	
	\begin{figure}[htbp]
		\centering
		\includegraphics[width=0.95\linewidth]{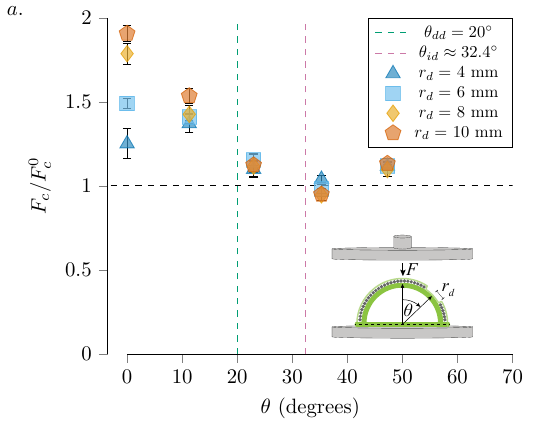}
		\includegraphics[width=0.95\linewidth]{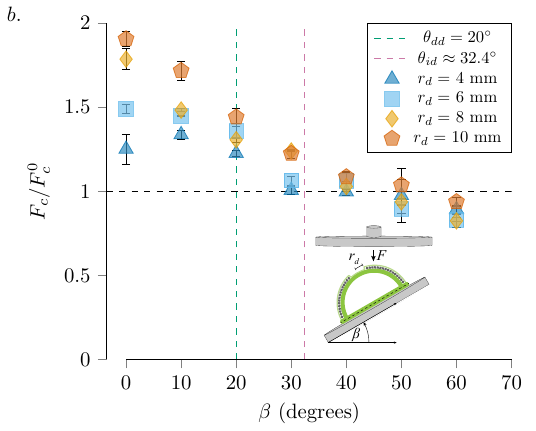}
		\caption{
			(a) The EG shell exhibits distinct normalized critical buckling loads $F_c/F_c^0$ at a defect angle $\theta=0^\circ$ but collapses to similar values as $\theta$ increases. $F_c^0$ refers to the critical buckling load for a shell without a defect and is the same for different values of $\theta$. The experimental setup for changing the angle of the defect $\theta$ is shown inset. 
			(b) As the angle of indentation $\beta$ increases, $F_c/F_c^0$ decreases and collapses to a similar value for the shell with different sized defects. $F_c^0$ refers to the critical buckling load for a shell without a defect, and is unique to each value pf $\beta$. The experimental setup for changing the angle of indentation $\beta$ is shown inset.
		} \label{defectFc}
	\end{figure}
	
	In Fig.~\ref{defectFc}a, we plot the critical buckling force of an EG shell with a defect with the critical buckling force of a shell without a defect as $F_c/F^0_c$ against $\theta$, which is the angle of the defect measured from the apex of the shell to the center of the defect. As we increase $\theta$, we observe a decrease in normalized critical buckling load until it reaches the buckling load of an EG shell with no defect (the dashed horizontal line) at around 35$^\circ$. The increase in critical buckling at $\theta \approx 47^\circ$ may be due to the defect interacting with the edge of the shell. 
	
	We denote angle thresholds for defect-defect and indenter-defect interactions based on previous literature with two dashed vertical lines as $\theta_{dd}$ and $\theta_{id}$, respectively. Both of these values refer to the interior angle between the center of one Gaussian dimple defect and another Gaussian dimple defect or point indenter on a thin elastic monolayer shell. In the work by Derveni et al. when two defects are within $\theta_{dd} = 20^\circ$ of each other, the interaction between the defects results in a decrease in knockdown factor; when the defects are greater than $20^\circ$ apart, the knockdown factor trends to a constant value coinciding with an equivalent one-defect case~\citep{derveni_probabilistic_2023}. For the case of indenter-defect interactions, Abbasi et al. derived the characteristic length of indentation $r_z=2.25\sqrt{Rh}$, which refers to the radius from the point indenter of the neighborhood within which the incidence of a defect would decrease the critical buckling load~\citep{abbasi_probing_2021}. For our EG shells, we set the thickness of the shell as $h_i$ to yield $r_z=16.13$ mm, which corresponds to $\theta_{id}=32.4^\circ$. Physically, this suggests that if the indentation and defect location are within 32.4$^\circ$, $F_c$ should decrease. However, we observed the opposite effect in EG shells: $F_c$ increases when $\theta<\theta_{id}$. Previous work has observed and discussed the issue of edge interactions if defects or indentation are close to the clamped boundary of a hemispherical shell and limited placing defects to within ~60$^\circ$ of the apex ~\citep{almroth_influence_1966,lee_geometric_2016,derveni_probabilistic_2023}. As our defects approach within 15\% of this boundary, edge effects may play a role in increased buckling loads. 
	
	Next, we explore the effect of eccentric loading.The defect is made at the apex of the shell ($\theta = 0$) and the shell is indented at an angle $\beta$, which measures the angle the indenter makes relative to the apex of the shell. Fig.~\ref{defectFc}b shows the results of measuring $F_c/F_c^0$ -- where each $F_c$ is now normalized by an $F_c^0$ corresponding to the same value of $\beta$ -- as the shell is compressed at different values of $\beta$. Once again, $\theta_{dd}$ and $\theta_{id}$ are denoted on the plot. As $\beta$ increases, critical buckling decreases for all defect sizes. At $\beta = 40^\circ$, $F_c/F_c^0$ collapses to a similar value regardless of defect size $r_d$, approaching the critical buckling load of an EG shell without a defect. Past $\beta=40^\circ$, dimensionless force drops below that of a defect-free EG shell.
	
	The results of varying both $\theta$ and $\beta$ have yielded similar findings. When indentation location and defect location are both axisymmetric to the apex of the shell, $F_c$ increases with $r_d$ due to increased $\phi(r_d)$. In both cases, as the angular distance the between indentation location and the defect location increases, the critical buckling force decreases and collapses to a similar value regardless of defect size. Although $\phi(r_d)$ may be higher due to a larger defect size, it is also serves as discontinuity in the granular media which can hinder force chain development when not masked by a axisymmetric indentation. This suggests that the effect that a defect has on critical buckling load is a local phenomena, with the boundary at $\theta_{id}\approx32.4^\circ$ for our tested EG shell. When the defect location and indentation location are less than $\theta_{id}$, the defect increases the critical buckling load, with larger defect sizes more strongly ``strengthening'' the EG shell. When the defect and indentation locations are greater than $\theta_{id}$ apart, their strengthening effect diminishes regardless of defect size, and show similar critical buckling loads to an EG shell without a defect.

	\subsection{Shape-Locking}
	
	When the inter-shell cavity is at ambient pressure, the EG shells can be easily deformed via indentation as the grains are in an unjammed, fluid-like state. Removing that indentation will result in the EG shell reverting to its original, undeformed hemispherical configuration. However, if the inter-shell cavity is subject to a negative pressure while an indentation is being applied, the shell will able to preserve the shape of the indentation. We can leverage this behavior to explore shape locking in EG shells. As the grains are also not longer arranged in a hemispherical shape upon indentation, the effect on shell stiffness is also investigated. 
	
	\begin{figure}[htbp]
		\centering
		\includegraphics[width=0.95\linewidth]{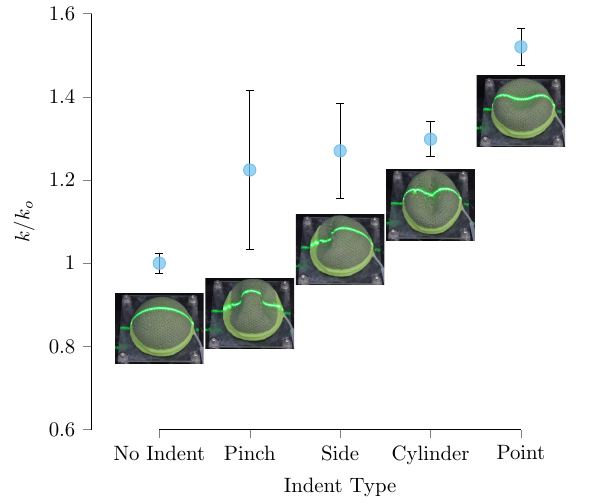}
		\caption{
			The normalized stiffness of the EG shells with various indentation types were compared to the an EG shell without an indent. The stiffness was measured at $\delta/R = 0.1$, consistent with earlier EG shells.
		}
		\label{shapeLocking}
	\end{figure}
	
	We tested four different indentation types using a slightly different EG shell than the ones used previously. These EG shells use a thinner outer shell of $h_o = 0.3$ mm so that the shell better conforms to the grains when subject to a negative pressure. The more closely the outer shell can conform to the grains, the more effective the shape locking effect is. The indentations applied to the EG shells are as follows: single indent at apex, single indent at the side, a cylindrical indent, and a "pinch" indent. An unindented shell was used as a comparison. A green laser is used to better show the geometry of the shells in different configurations. The stiffness response at $p_i = 25 kPa$ and images of these shells are shown in Fig.~\ref{shapeLocking}. Shells were only loaded within the linear regime. We see that the point indent caused the greatest increase in stiffness. This may be due to the symmetric shape that allows the force chains to be evenly distributed around the ridge of the indent. No indented shape was weaker than a hemispherical shell, suggesting perhaps that a higher, local, positive Gaussian curvature near the indenter allows for a higher stiffness response. This idea comes from previous work by Vella et al. and Lazarus et al. which found that the indentation force response to displacement $\delta$ is $F = k_1 \delta$ where~\citep{vella_indentation_2011,lazarus_geometry-induced_2012}
	\begin{equation}
		k_1 = 8\sqrt{BEh\kappa_G}
	\end{equation}
	with bending stiffness $B$, shell thickness $h$, and Gaussian curvature $\kappa_G$. Stiffness $k_1$ scales with $\kappa_G$ and therefore, similar structures with higher Gaussian curvature should exhibit a higher force response.
	
	\section{Conclusion}
	We have introduced a new multilayer elastogranular shell structure that exhibits tunable stiffness and buckling properties. By changing the packing fraction of the grains during EG shell construction and controlling the pneumatic pressure in the inter-shell cavity, we were able to achieve a tunable stiffness up to 3.5 times the unactuated EG shell. We built upon classical shell theory to empirically derive a new effective compressive stiffness for elastogranular shells. In the large deflection regime, we suggest that the critical buckling load is determined by the presence of vacancy defects in the granular media. When we intentionally introduce a vacancy defect and control its size, we observed that critical buckling load increases with defect size. We were able to confirm this by using our scaling relation to predict critical buckling load from the defect size and initial packing fraction, for which we saw good agreement. We found that when the indentation location and the defect location are within $\theta_{id}$ of each other, there is a ``masking effect'' that reinforces the EG shell and allows it to buckle at a higher applied load. This effect weakens as the indentation location and the defect location are more than $\theta_{id}$ apart. Finally, we utilized the compliance switching ability of the EG shell to explore shape-locking behavior and found that the higher Gaussian curvatures presented by indented shapes may be the reason for increased stiffness behavior. 
	
	We believe shells that shape-shift and compliance-shift can emerge just by changing the sizes and stiffnesses of the shells and confined grains relative to each other. Creating an EG shell with stiff inner and outer shells coupled with soft grains may allow for the grains to effectively increase their packing fraction as they are compressed between the shells. Grains much smaller than the thickness of the shells may allow the structure to be manipulated into even more complex geometries for shape-locking behavior. A thin inner shell and thick outer shell could be a method for selectively reinforcing a smooth load-bearing hemispherical structure. Explicitly measuring curvature of the indented shapes in the EG shells could allow us to compare the stiffness scaling based on curvature of our EG shells to existing theory on ellipsoidal shells.
	
	Buckling behavior is an important characteristic of shells that is still extensively studied to this day~\citep{zoelly_ueber_1915,hutchinson_imperfection_1967,lee_geometric_2016,derveni_most_2025}. Advances in fabrication techniques have allowed for shells that can achieve higher knockdown factors before collapse, but outside of a few works~\citep{yan_magneto-active_2021,lee_evolution_2019}, the buckling behavior of these structures are set during fabrication. Through the use of differential internal pressure and vacancy defects, we have been able to achieve tunable buckling properties after structure fabrication. We believe that this work on EG shell structures contributes to the recent trend towards metamaterials with tunable and reversible physical properties which motivate the design of more efficient structures.
	
	\section{Appendix}
	
	\subsection{Elastogranular Shell Fabrication}
	
	Three constituent parts are assembled together to create an EG shell: a thick elastic hemispherical bottom shell, a thin elastic hemispherical top shell, and a single layer of spherical rigid grains which are encapsulated by the two shells. The top and bottom shell fabrication methods are based on the work by Lee et al.~\citep{lee_fabrication_2016}. A smooth metal sphere (radii $R_m \in$ [28.58, 31.75, 34.91] mm, McMaster-Carr) serves as the mold and is coated with mold release (Mann Release Technologies Ease Release 200) and then coated with polyvinyl siloxane (PVS, Elite Double, Zhermack) and let cure for 20 minutes at room temperature. We used three different PVS formulations: Elite Double 8, Elite Double 32, and Elite Double 32 Fast, which had measured Young's moduli of 0.23 MPa, 0.95 MPa, and 0.89 MPa, respectively. We measured the Young's modulus via uniaxial tension tests on a dogbone-shaped sample. We note that PVS is commonly approximated as incompressible, {\em i.e.} it has a Poisson's ratio $\nu \approx$ 0.5~\citep{marthelot_buckling_2017}. Thicker shells are fabricated by a repeated layer-by-layer deposition of PVS, {\em i.e.}. the process of coating and curing PVS on the same sphere is repeated until the desired thickness is achieved. After achieving the desired thickness, the film is peeled off to yield an approximately hemispherical elastic shell. The outer shell has a thickness $h_o = 0.6$ mm and various inner shell thickness were tested (inner shell thickness $h_i \in$ [1.5, 1.8, 2.1] mm). Shells of different radii were constructed to vary $h_i/R$,
	
	\begin{figure}[htbp]
		\centering
		\includegraphics[width=0.95\linewidth]{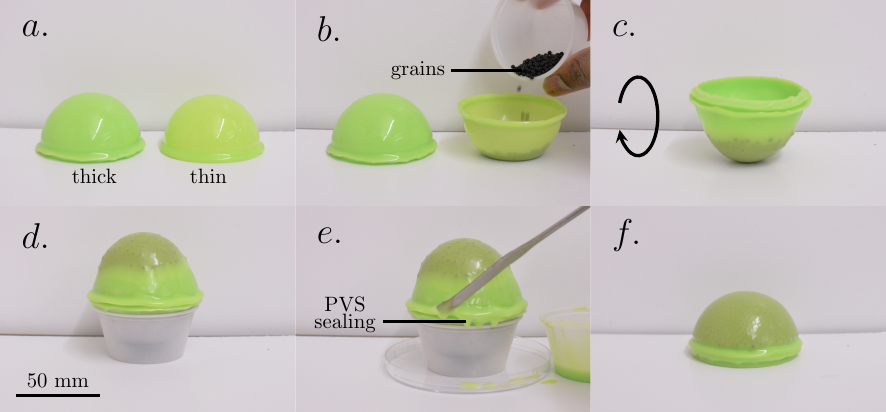}
		\caption{ 
			The different stages of EG shell construction in order. (a) The outer and inner shell are made with multiple layers of PVS as in~\citep{lee_fabrication_2016}. (b) Grains are poured into the concave side of the outer shell. (c) the convex side of the inner shell is placed on the concave side of the outer shell to sandwich the grains. (d) The assembled structure is placed on a metal sphere to allow for further fabrication. (e) More liquid PVS is used to seal the grains between the two shells. (f) the PVS is allowed to dry and excess is cut off to yield the finished EG shell. Additionally, an incision as made at the edge of the shell so that the a pressure line can be attached for pneumatic control.
		}
		\label{shellFab}
	\end{figure}
	
	\begin{figure}
		\centering	
		\includegraphics[width=0.95\linewidth]{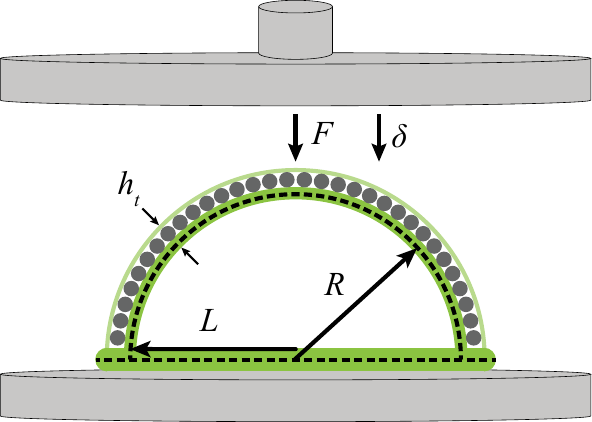}
		\caption{
			The indentation is axisymmetric to the apex of the shell while force $F$ and displacement $\delta$ are measured.
		}
		\label{shellParameters}
	\end{figure}
	
	The elastogranular mechanics that we studied in this work consist of nearly hemispherical shells, with a span L over radius R of the different sized shells range between 0.9 $< L/R <$ 1. After the top and bottom shells are fabricated, spherical glass beads (radii $r_g \in$ [1, 1.3] mm radius,  mass $m_g \in$ [0.01, 0.02] g), Jan's Jewelry Supplies) are sandwiched between the concave side of the outer shell and the convex side of the inner shell and sealed with PVS. The region between the shells that contains the grains will be referred to as the inter-shell cavity. An incision is made at the rim of the EG shell between inner and outer shell so that a plastic tube may be inserted to allow for pressure control of the EG shell. Images of the fabrication steps are also shown in~Fig.~\ref{shellFab}.
	
	Conventionally, the radius of a shell is measured to its middle surface, which would be at half the thickness of a homogeneous, single layered shell. Although EG shells have multiple layers of multiple thicknesses, we will still consider the middle surface as the radius to half the thickness of the inner shell. We defined this radius to the middle surface as $R = R_m + h_i/2$. Schematics of the various relevant EG shell parameters are shown in~Fig.~\ref{shellParameters}. The location of our proposed middle surface is marked with the dashed line.
	
	The thin outer shell is compliant, and so the grains can be manually spread into a thin monolayer in the inter-shell cavity by externally nudging them. We defined $\phi_0$ as the granular packing fraction when the shell is at $p_i = $ 0 kPa. We can experimentally vary $\phi_0$ by changing the number of enclosed grains during EG shell construction. On a flat surface, spheres in crystalline order exhibit a maximal hexagonal packing fraction of $\phi_0^{\max}$ = 0.91~\citep{chang_simple_2010}. However, this crystalline order is geometrically incompatible with a curved surface and therefore requires a number of defects -- which in this context refers to point defects and line defects in the lattice -- to accommodate the curvature~\citep{seung_defects_1988}. In packing spherical grains on a curved surface in our system, we used a max packing fraction $\phi_0^{\max}$ = 0.86~\citep{schreiner_computer_1982,mascioli_defect_2017}. This value is derived from previous work on packing disks and spheres on a spherical surface, in which the number of elements being packed determines the specific maximum packing density achievable. We calculated $\phi_0$ for our system as
	\begin{equation}
		\label{phi0calc}
		\phi_0 = \phi_0^{\max}\left(\frac{n_g A_g}{A_s}\right) 
	\end{equation}
	where $n_g$ is the number of grains in the EG shell, $A_g = \pi r_g^2$ is the cross sectional area of a singular grain, and $A_s = 2 \pi (R + h_i/2 + r_g/2)^2$ is the surface area of the spherical cap at the centerline of the monolayer of grains. EG shells of various packing fractions ($\phi_0 \in$ [0.45, 0.5, 0.58, 0.66, 0.71, 0.75, 0.8, 0.85]) were fabricated and tested in this work.
	
	\subsection{Compression Tests}
	
	An EG shell is placed on an acrylic plate which is screwed onto a multi-angle mount (410 Junior Geared Head, Manfrotto). The tubing on the EG shell is connected to a MPRLS Ported Pressure Sensor ($0-172.37$ kPa, Adafruit Industries) which is interfaced with an Arduino to monitor the pressure inside the shell. It is also connected to a syringe pump (Harvard Apparatus) to allow for manual pneumatic control of the EG shell. 
	
	The inter-shell cavity is pressurized using the syringe pump. The differential internal pressure $p_i$ is measured as the magnitude of the difference between the air pressure inside the EG shell and the ambient air pressure outside the EG shell. For example, $p_i$ = 10 kPa denotes that the pressure in the inter-shell cavity is 10 kPa lower than the atmospheric pressure. For all the experiments, the EG shell is held at a constant level of differential internal pressure ($p_i \in$ [$0, 5, 10, 15, 20, 25$] kPa). Pressures greater than 25 kPa were not tested because they resulted in tearing in the outer shell.
	
	Compressive indentation testing was conducted using a mechanical testing machine (Instron 5943) with a 500 N load cell. The experiments yield force ($F$) vs indentation distance ($\delta$), or compression, data which is then analyzed for specific shell characteristics. The shell is indented using a large, flat, circular acrylic plate with a radius of 50 mm. A thin sheet of polytetrafluoroethylene (PTFE) is adhered to the indenter surface and talc powder is spread on the EG shell surface before testing to reduce frictional effects between the plate and the EG shell. The talc is reapplied every three trials to minimize friction. The indenter velocity during indentation tests is set to 10 mm/min to ensure that the mechanical response is effectively quasi-static. The displacement of the indenter is increased from $\delta$ = 0 at the apex of the shell to $\delta \approx 0.5R$. A schematic of the experimental setup when testing the effects of internal pressure and grain density is shown in~Fig.~\ref{shellParameters}. The experimental setup for testing defect location is shown in~Fig.~\ref{defectFc} a inset. The indenter angle $\beta$ is the angle that the surface that the shell rests on is tilted with respect to its starting flat position ($\beta = 0^\circ$) and was tested between $0^\circ \leq \beta \leq 30^\circ$. Unless otherwise specified, $\beta =$ 0 which signifies that compression is axisymmetric to the apex of the shell. A schematic of the experimental setup for testing the angle of indentation is shown in~Fig.~\ref{defectFc} b inset.
	
	\subsection{Calculation and Visualization of Force Chain Development}
	\label{simulationsInfo}
	For the simulations done in Fig.~\ref{forceChainMechanism}b, we approximated the grains as disks with the same radius $r_g$. The disks are randomly placed without overlap on a spherical dome of radius $R + h_i/2 + r_g/2$ - which is the imaginary spherical cap cutting through the center of the monolayer of grains on top of the inner shell. Initially, the disks are evenly spaced on the spherical dome, which gives them a spacing $s_g$, or distance between the centers of two neighboring disks, based on the packing fraction. Therefore, the spacing between the edge of two neighboring disks is given as $s_g-d_g$ where $d=2r_g$ is the diameter of the disk. Experimentally, however, the grains in the EG shell are packed with dispersed empty regions as well as regions of hexagonal order, as seen in Fig.~\ref{shellOverviewPics}b. To average the different granular configurations, each disk on the dome is perturbed in a random direction from 0 to 2$\pi$ and a random distance from 0 to $(s_g-d_g)/2$. In this new configuration, if two disks are within a grain separation threshold distance $s_{gT}$ -- calculated from previous work by Schunter et al.~\citep{schunter_elastogranular_2018} and is graphically presented in Fig.~\ref{conformalityExp}a -- they are considered to be in contact. To be counted as a complete force chain, there must be an arrangement of consecutive disks within $s_{gT}$ of each other starting from a disk within 5$^\circ$ of the apex of the dome to one of the disks at the edge of the dome. Each subsequent disk in a force chain is also limited to moving down the dome's meridian.
	
	There are two simulation programs: one calculates the force chain success rate of a range of $\phi_0$ values while the other visualizes the force chains on a spherical dome. The calculation script takes in inputs for EG shell parameters (spherical cap diameter, spherical cap height, and grain diameter) and tests it over the desired starting and ending values of $\phi_0$ as well as number of trials. It outputs a plot of the force chain success rate -- defined as the number of trials out of total for which a force chain was detected, for the range of $\phi_0$ tested. The visualization script takes in the same EG shell parameters and a value of $\phi_0$. The program outputs the force chain success rate, number of force chains found, and a visualization of the spherical dome with the force chains highlighted, if there are any. The code for both of these simulations are provided at GitHub~\citep{narendran_elastogranularshells_2026}.
	
	\subsection{Grain Conformality}
	
\begin{figure}[htbp]
		\centering
		\includegraphics[width=0.95\linewidth]{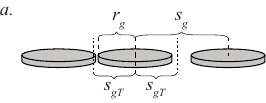}
		\includegraphics[width=0.95\linewidth]{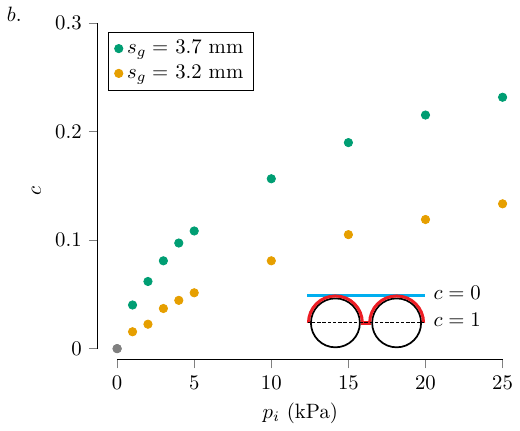}
		\caption{ 
			(a) Modeling the grains as disks for the simulation work done in Fig.~\ref{forceChainMechanism}. The radius of the disks are given by $r_g$, the separation distance between two grains is given by $s_g$, and the threshold distance between two grains to allow force chain propagation is given by $s_{gT}$. (b) In this simplified 2D model, two values of grain separation distance $s_g$ are tested for their conformality when $p_i$ is increased. For both of these cases, as $p_i$ increases, conformality c increases as well. This measure of conformality is taken from~\citep{guerra_elastogranular_2023}.
		}
		\label{conformalityExp}
	\end{figure}

	Guerra et al. defined conformality $c$ as how much a thin sheet is wrapped around an individual grain, with $c = 0$ denoting no conformality and $c = 1$ denoting that half of the grain is conformed to~\citep{guerra_elastogranular_2023}. We made a simplified 2D model of the outer layer and grains to understand conformality in our EG shell system. In~Fig.~\ref{conformalityExp} b, we plot how conformality changes with $p_i$ for two grain distances $s_g$ which encapsulates the average grain distances in our EG shells. At $p_i =$ 0 kPa, the conformality is 0 and the outer shell does not conform to the grains. As $p_i$ increases, the conformality $c$ also increases, denoting that the thin outer layer is more tightly conformed to the grains. The $s_g = 3.2$ mm corresponds to grains that are touching (expected for shells with $\phi_0 \approxeq \phi_0^{max}$) while $s_g = 3.7$ mm corresponds to grains that are 0.5 mm apart (expected for shells with $\phi_0 \approxeq 0.44$).
	
	\subsection{Creating Packing Defects}
	
\begin{figure}[htbp]
		\centering
		\includegraphics[width=0.95\linewidth]{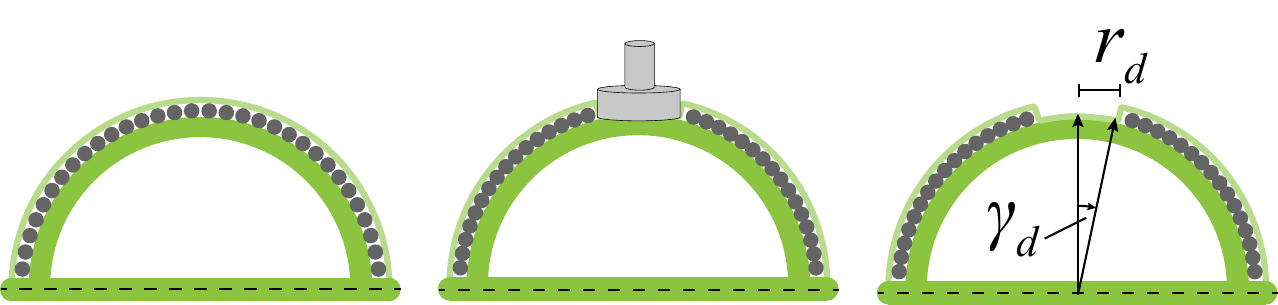}
		\caption{ 
			The creation of an EG shell with a defect: First the inter-shell cavity is set to $p_i$ = 0 kPa. Then a circular acrylic disk is pressed onto the EG shell and the grains are rearranged around the disk. The air is evacuated from the inter-shell cavity and the acrylic disk is removed to yield a shell with a vacancy defect of radius $r_d$.
		}
		\label{packingDefectCreation}
	\end{figure}

	Granular packing defects are introduced to EG shells by creating a circular region on the shell that is devoid of any grains.  The inter-shell cavity is subject to an ambient pressure, which allows the grains to move freely. The grains can be physically rearranged around a circular acrylic disk that is pressed on the shell at the desired location for the defect. This moves the grains out of the desired area of the defect and creates an empty void in the granular arrangement. The size of the acrylic disk (radii $r_d \in$ [$4, 6, 8, 10$] mm) will also be the size of the defect. When the air is evacuated from the EG shell, the outer elastic shell conforms to the grains. The acrylic disk is now removed and the defect will keep its shape (\ref{packingDefectCreation}). If the defect is not at the apex of the shell, $\theta$ is also used to measure the angular position of the defect with respect to the apex (e.g. as depicted in Fig.~\ref{defectFc} a). Returning the shell back to ambient pressure relaxes the confinement of the grains and allows them to move freely again. This reversible technique of defect creation allows the testing of multiple defect sizes and locations on a single EG shell.
	
	\subsection{Indentation for Shape-Locking}
	
\begin{figure}[htbp]
		\centering
		\includegraphics[width=0.95\linewidth]{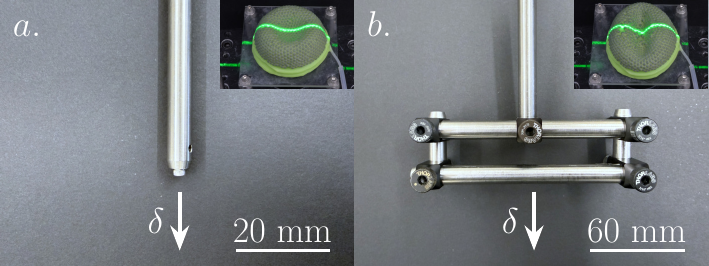}
		\caption{ 
			(a) A point indenter using a screw of radius 2.58 mm. (b) A cylindrical indenter using a cylindrical rod of radius 6.4 mm. The indentation direction for shape locking is given by $\delta$.
		}
		\label{indenterImage}
	\end{figure}

	When studying shape-locking effects of an EG shell, an indentation had to be locked into the EG shell before compression tests could be performed. We used an EG shell with $h_o = 0.3$ mm to get better conformation of the shell to the grains. The EG shell is kept at ambient pressure and an indentation type is applied to the shell. The head of a screw with a radius of 2.58 mm was used as the point indenter, a cylindrical rod of radius 6.4 mm was used as the cylindrical indent, and manual indentation using fingers was used for the side and pinch indent. Images of the point and cylindrical indenter are shown in Fig.~\ref{indenterImage} with the resulting indentation shape shown inset. Air is evacuated from the inter-shell cavity while the indentation is held. When $p_i = 25$ kPa, the indentation can be removed and the EG shell with hold the shape of the indentation applied. The EG shell can be tested to determine its stiffness characteristics as described in Appendix A.2.
	
	\section{Acknowledgments}
	The authors gratefully acknowledge the financial support from NSF CMMI-MOMS (\#2414101).
	
	\section{Data and code availability}
	The code required to run these simulations is available at the following github: https://github.com/MoSS-BU/ElastogranularShells.

	\bibliographystyle{rsc}

	\bibliography{EG_Shell_Bib}

	\end{document}